\documentclass{evn2004}
\usepackage{txfonts}
\begin{document}
\setcounter{page}{303}

   \title{Applications of precision astrometry to studies of massive YSOs}

   \author{Mar\'{\i}a J. Rioja\inst{1}
          \and
          Luca Moscadelli \inst{2}
          \and
          Riccardo Cesaroni \inst{3}
          }

   \institute{Observatorio Astron\'omico Nacional, Apdo. 112, 28803 Alcal\'a
              de Henares (Madrid), Spain
         \and
             Osservatorio Astronomico di Cagliari, Loc. Poggio dei Pini, 
             strada 54, 09012 Capoterra (Cagliari), Italy
         \and
             Osservatorio di Arcetri, Largo E. Fermi 5, 50125 Firenze, Italy
             }
   \abstract{
We present VLBI observations of the H$_2$O maser emission towards a 
selection of massive young stellar objects (YSOs).
In one of these sources, IRAS20126+4104, the measured proper motions confirm
that the H$_2$O masers spots are tracing the very root ($\sim$ 100 AU)
of a conical bipolar outflow emerging near the position of the embedded YSO,
and are an excellent tool to investigate 
the structure and kinematics of the outflow/jet.
We also present preliminary results of a first epoch EVN observations 
of a selection of 5 high-mass YSO to assess the precise association of 
maser spots and molecular outflows in these sources.
   }

   \maketitle
%

\section{Introduction}

\noindent
There is still much to learn about the role played by molecular outflows 
in the context of star formation processes.
This phenomenon presents many interesting aspects
ranging from the possible role in supporting molecular clouds against
gravitational collapse, to the collimation mechanism, to the velocity
structure of the outflows on different scales. From an observational 
point of view, answering these questions is equivalent to attaining the best
possible picture of an outflow in space and velocity. Since the outflowing
material is mostly molecular, normally the means used to image these 
objects are the transitions of molecules such as CO, HCO$^+$, and a few
others (see e.g. Bachiller and P\'erez Guti\'errez, 1997),
which usually lie in the millimeter range. However, these tracers
can be used only to map the large scale structure of the flow (0.1-1 pc)
and not the very root of it, which is likely to convey the most important
information on the ejection mechanism, being very close to the central 
engine.  Such a region has a size of a few 10 AU, not accessible even with
the most powerful millimeter interferometers, but easy to study with VLBI
techniques at centimeter wavelengths. The molecular transitions commonly used
to map outflows lie in the millimeter range; however, molecular outflows
from high-mass YSO are often associated with H$_2$O maser emission at 22 GHz
(Felli et al. 1992), which hence represents the best tracer for studies of the
flow at a small scale. Water masers have very large brightness temperatures,
which make them ideal targets for VLBI studies. \\

With this in mind we have carried out VLBI observations of H$_2$O 
masers in a selection of high-mass young stellar objects which belong to 
the list studied by Tofani et al. (\cite{tofani95}) with the VLA in the 
most extended configuration, to take one step further in the analysis of 
spatial distribution and kinematics of the masers. 
The case of IRAS20126+4104 deserves special mention due to the numerous
studies of different molecular tracers on scales ranging 
from 100 AU to 1 pc
(Cesaroni et al. 1997; Zhang et al. 1998; Cesaroni et al. 1999a; Hofner et al.
1999; Zhang et al. 1999) 
that resulted in the detection of a rotating Keplerian 
disk around the YSO and a detailed analysis of the jet/outflow, with the 
large scale outflow being fed by a narrow jet, which is ionised on a scale 
of $\sim 1''$ (1700 AU) and becomes neutral on $\sim 20''$ (0.16pc).
IRAS20126+4104 is considered to be the most convincing case of a disk-outflow 
system in a massive YSO. Moreover, our VLBI observations of H$_2$O 
maser emission (Moscadelli et al. 2000) showed a very good agreement 
with a jet model 
which assumes that the masers arise on the surface of a conical bipolar jet, 
at the interaction zone between the ionised jet and the surrounding neutral 
medium. Hence the spots are tracing the very root ($\sim$ 50 AU) of a 
bipolar outflow.
Follow up proper motion measurements of the H$_2$O maser spots with 
new multi-epoch VLBI observations have
proved the uniqueness of this interpretation. Throughout this paper we 
assume a distance to IRAS20126+4104 equal to 1.7 kpc. \\

The encouraging results found for IRAS20126+4104 provided the ground for 
more VLBI observations towards other sources from the Tofani et al. 
(\cite{tofani95}) sample, aiming 
to assess the precise association of maser spots and molecular outflows.
Table ~1 lists the selected sources on the basis of the following criteria:
(i)~to be deeply embedded in dense molecular clumps; (ii)~to be undetected 
in the free-free radio continuum (i.e. {\it not} associated with HII regions);
(iii)~to have luminosities above 1000~$L_\odot$;
(iv)~to lie close to a compact (in VLBI scales) continuum reference source
(for observations in phase referencing mode).
The third condition guarantees that one is dealing with high-mass YSOs,while
the first two bias the sample towards the youngest, least evolved star-forming
regions. For all selected sources Cesaroni et al. (\cite{cesa99b}) detected 
high density molecular clumps around the H$_2$O masers.

\section{Observations and data reduction}

We carried out VLBA 22 GHz observations of the
young massive (proto)star IRAS20126+4104 on November 21-22, 1997,for a total
of 12 hours.
The analysis of these observations proved the association of the maser
spots with the bipolar outflow and provided an estimate of the velocity
field on a scale of $\sim 100$ AU. A complete description can be found 
in Moscadelli et al. (\cite{mcr}). \\
The study of the source was followed up with 3 
VLBA+EVN (global) astrometric observations to measure H$_2$O maser spot 
proper motions. The multi epoch observations were scheduled in the course of
4 months to match the mean lifetime of maser lines, estimated on the
basis of our single dish monitoring. 
The global campaign included all VLBA, and 8 EVN antennas:
Effelsberg (100m, Germany), Medicina (32m, Italy), Noto (32m, Italy), 
Onsala (20m, Sweden), Jodrell Bank (25m, UK), Metsahovi (14,Finland), Sheshan 
(25m, China) and DSS63 (70m, Spain)
observing  on Nov. 9th and 26th, 2000, and March 1, 2001, for a total 
of 18 hours, at 22 GHz. During the observations, the antennas switched 
every 30 s between IRAS20126+4104 and the continuum calibrator source 
J2007+4029, $1^0$.5 apart. The quasar J2007+4029 
belongs to the list of sources used to define the International Celestial 
Reference Frame (ICRF) and has very accurate coordinates (uncertainties 
in RA and DEC $< 1$ mas). 
All stations recorded an aggregate of 16 MHz bandwidth in each (left and 
right circular) polarization for each scan, centered at the LSR velocity 
of -3.5 km s$^{-1}$ (based upon a rest frequency of 22235.0798 MHz), 
using 2-bit sampling (mode 128-2-2). 
The correlation was made at the VLBA correlator in Socorro (New Mexico) 
using 1024 spectral points which led to a channel separation of 0.21 
km s$^{-1}$. \\

\begin{figure}
   \centering
   \vspace{250pt}
 \includegraphics{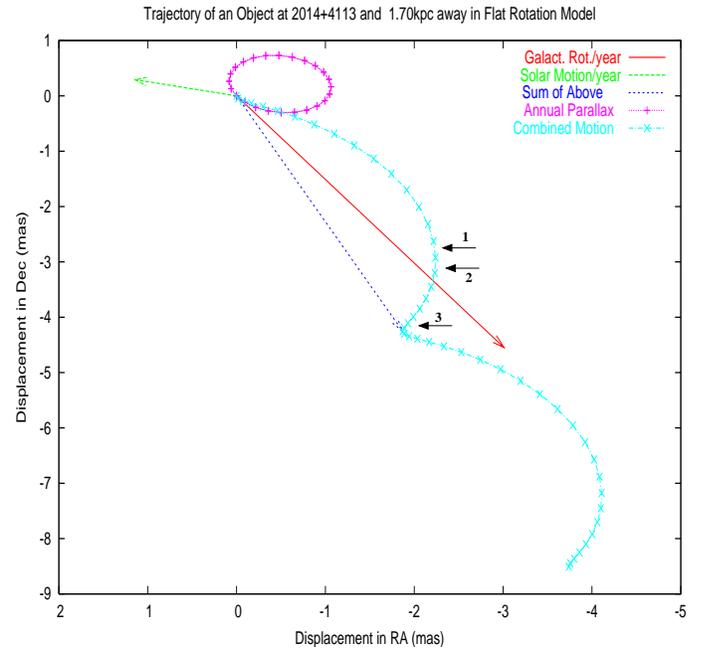}
      \caption{Simulation of the trajectory of an object 1.7 kpc 
away due to the Galactic rotation (solid line), the annual parallax (dotted
line) and the solar motion (long dashes line). The Galactic rotation is 
assumed to follow the simple Flat Rotation Model(FRM).
The combined motion follows the ``x'' line, for a time span of 2 year. 
The VLBI multi-epoch observing dates are indicated with labels 1,2, and 3, 
corresponding to first, second and third campaigns, respectively. 
          \label{fig:galrot}
                }
   \end{figure}

The data reduction was done using the NRAO AIPS software package.
The information on system temperatures (T$_{sys}$), gain curves 
and telescope gains measured at the individual array elements was used 
to calibrate the raw correlation coefficients of the line and reference 
sources. The application of standard fringe-fitting, amplitude and phase 
(self-)calibration techniques produced a hybrid map of the reference 
source. \\
The phase calibration of the line source involved a temporal 
interpolation between adjacent scans on the reference source. 
This strategy preserves the signature of the relative separation, 
between the target and reference source pair, present in the calibrated phase. 
In our series of observations the rapid antenna switching matched the 
requirements for a succesfull astrometric analysis and
the Fourier Transformation of the calibrated visibility function of the line 
source produced multiple ``phase referenced'' maps, corresponding to 
different spectral channels with line emission. At each epoch, the offsets 
of the maser spots from the center of the map are estimates 
of the absolute position parameters in the astrometric analysis. 
Moreover, a multi epoch comparison (at least 3 different epochs
in the course of $\le 1$ year) leads to proper motion estimates.
Fig. ~\ref{fig:galrot} shows the displacements, in RA and DEC coordinates, 
for a galactic source 1.7 kpc away  due to the 
Galactic rotation, the annual parallax and the solar motion, for 
a time span equal to 2 years. 
We have implemented this ``galactic motion'' effect in the calculus of the
maser proper motions. A more detailed description of the analysis of the
multi epoch observations will be given in Moscadelli et al. 
(\cite{mosca2004}).\\

%

\vspace*{-0.5cm}

Also, EVN observations of H$_2$O maser emission towards other 
5 high-mass young stellar objects, along with nearby ($< 3^o$ apart) 
continuum calibrator sources, were carried out on 
February 10-11, 2004. 
We observed with an array of 9 antennas  for a total duration of 15 hours, 
at 22 GHz.
The observing network was composed of the following antennas: 
Cambridge (32m, UK), Jodrell Bank (25m, UK), Effelsberg 
(100m, Germany), Noto (32m, Italy), 
Onsala (20m, Sweden), Metsahovi (14m, Finland), Sheshan (25m, China)
 plus 2 other antennas wich failed to produce interferometric fringes.
At the observations, all stations used a similar configuration as the one 
described above for IRAS20126+4104.
For a given pair of line/continuum calibrator sources the recorded 
bandwidth was centered at the LSR velocity of the line source (based upon 
a rest frequency of 22235.0798 MHz), derived 
from single dish observations obtained with the Medicina radiotelescope 
few days before the VLBI run.  
The processing of the data was done at the JIVE correlator, in Dwingeloo, using
1024 spectral channels. 
For most of the time the antennas were switching, with 3 minutes duty cycles,
between the 2 sources at each pair. 
This sequence was interrupted every 2 hours with 3 minutes
scans on the continuum calibrator sources 3C84 and 0528+134.
Our preliminary analysis of the EVN observations does not implement 
phase referencing techniques, and follows the standard procedure for the 
analysis of spectral line data within AIPS.

\section{The interesting case of IRAS20126+4104}

\begin{figure}
   \centering
   \vspace{300pt}
 \includegraphics{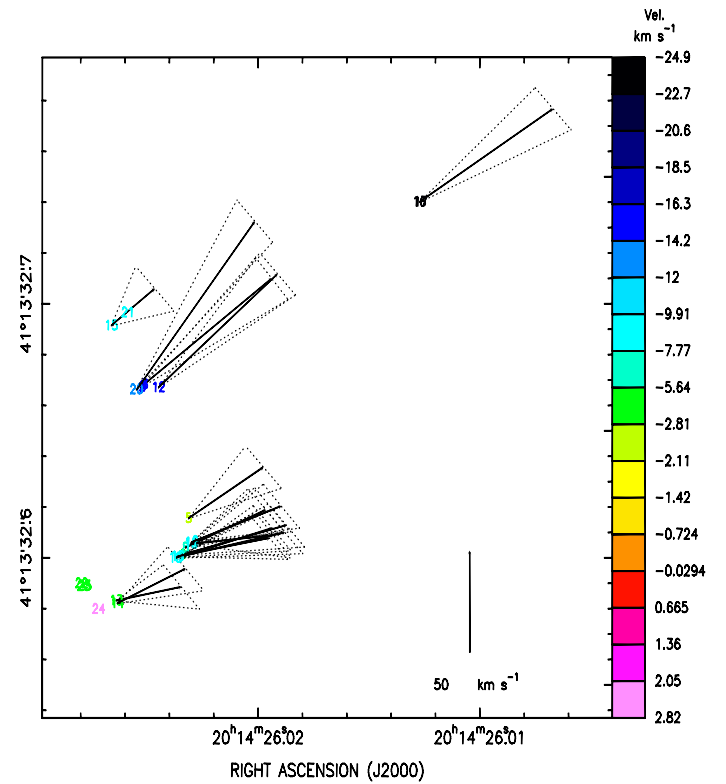}
      \caption{Map of H$_2$O maser spots and measured proper motions, measured
               from global VLBI multiepoch observations.
         \label{fig:propmot}
                }
   \end{figure}

\begin{figure}
   \centering
   \vspace{450pt}
\includegraphics{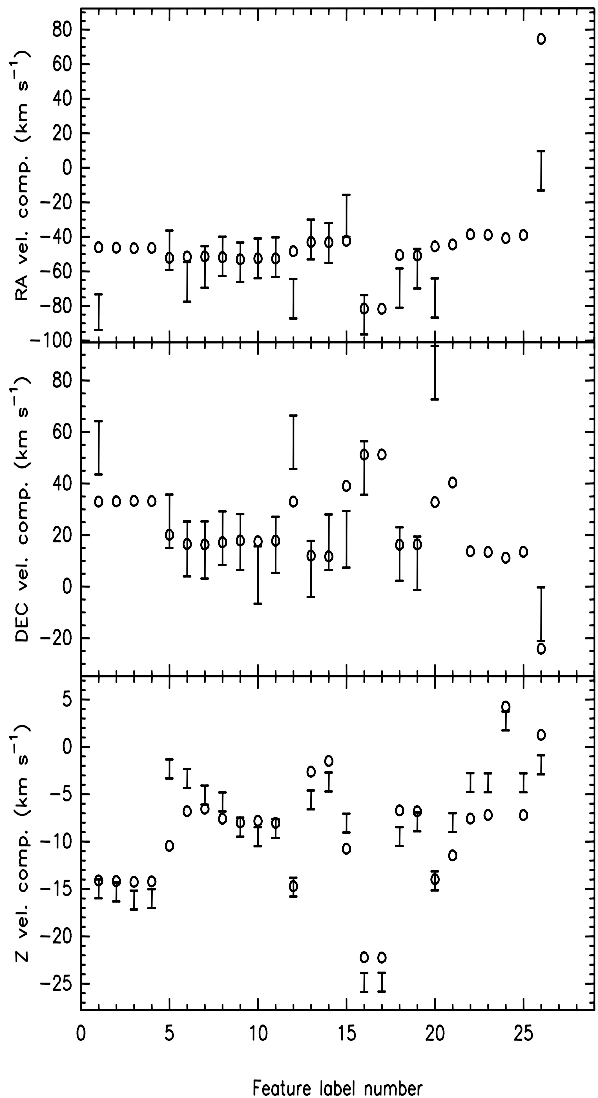}
      \caption{Comparison between the observed maser velocity components 
and those obtained from the best model fit, for data from global multiepoch
VLBI observations. 
         \label{fig:maservel}
                }
   \end{figure}

Our previous single epoch VLBA observations (Moscadelli et al. 2000)
detected 26 H$_2$O maser spots spread over a region of ~$\sim 0''.7$
(1200 AU).
The VLBI data are in excellent agreement with the predictions of a 
model which assumes that the H$_2$O masers lie on the surface of a conical 
bipolar jet, and expand radially away with constant velocity from a 
common centre coincident with the position of the YSO. Figure 5 in 
Moscadelli et al. (2000)
show a map with the spatial distribution of the H$_2$O maser spots and 
a comparison 
between the spot LSR velocities observed and computed from the 
best fit of the free parameters of the model (vertex position, opening angle 
of the cone, inclination of the cone axis with respect to the line of 
sight, and velocity of the spots).\\
Fig.~\ref{fig:propmot} shows the map of the H$_2$O maser spots for the new
multi epoch global observations, along with measured proper motions.
The best fit of model parameters to the data from multi epoch global 
observations is obtained for a well collimated
(semi opening angle=17$^0$) conical bipolar jet model 
with the vertex near the peak of the 3.6 cm and 7 mm continuum emission 
recently observed with the VLA (Hofner personal comm.),
and its axis closely aligned to the plane of the sky (the angle between the 
jet axis and the line of sight is 96$^0$) and with the direction of measured 
proper motions (position angle north-to-east=122$^0$). 
The fit of the multi epoch observations
implements a Hubble velocity outflow, with a variation of the velocity  
proportional to the distance to the vertex of the cone. 
Fig.~\ref{fig:maservel} shows a comparison between the observed maser 
velocity components and those obtained from the best model fit. 
Given the goodness of the fit, the new results prove the uniqueness 
of the interpretation proposed by Moscadelli et al. (\cite{mcr}), 
thus validating the use of H$_2$O masers as excellent tools 
to investigate the structure and kinematics of the jet/outflows 
in massive YSOs. \\
The combination of the proper motion measurements, from the astrometric 
analysis of the multi epoch observations, and the velocities along the 
line of sight, allows us to compute total velocities. We
stress that this is completely independent of the conical model.
The measured total velocities of the H$_2$O maser spots are in a range 
between 11 km ~s$^{-1}$ and 113 km ~s$^{-1}$, with most of them with 
velocities around 60 km ~s$^{-1}$.
This value is larger than our previous estimate (23 km ~s$^{-1}$), 
from the best fit of the single VLBA observations to the conical model
(Moscadelli et al. 2000). Interestingly, the new velocities are closer 
to the values estimated by Cesaroni et al. (1999a) from the SiO jet, 
between 60 and 200 km ~s$^{-1}$, on a much larger scale.

\vspace*{-0.4cm}
\def\kms {\rm ~km~s$^{-1}$}
\begin{table}[h]
\caption[ ]{ VLA H$_2$O Maser Properties of YSO observed with EVN.}
\label {vlatofani}
\begin{flushleft}
\begin{tabular}{lrrrrrrr}
\noalign{\hrule}
\noalign{\smallskip}
\hline
\noalign{\smallskip} 
Source & \multispan2 \hfil {Radio Coordinates (B1950)} \hfil & Peak Flux & Integ. Flux &
\multispan3 \hfil velocity (km s$^{-1}$) \hfil \cr
Component & R.A. & Decl. & (Jy) & (Jy km s$^{-1}$) & $v_{\rm peak}$ & $v_{\rm min}$ & $v_{\rm max}$ \\
\noalign{\smallskip}
\hline
\noalign{\smallskip}
NGC281-W & 00 49 28.233 & 56 17 26.408 & 9.38 $\pm$ 0.48 & 17.66 $\pm$ 0.46 & -27.9 & -39.8 & -25.9 \cr
S233 & 05 35 51.199 & 35 44 12.975 & 5.39 $\pm$ 0.27 & 12.29 $\pm$ 0.29 & -16.9 & -18.9 & -14.3 \cr
S235 B & 05 37 31.864 & 35 40 17.775 & 165.73 $\pm$ 8.29 & 257.51 $\pm$ 6.38 & -61.2 & -69.1 & -55.3 \cr
GGD 12-15 & 06 08 25.662 & -06 10 49.60 & 104.67 $\pm$ 5.24 & 240.48 $\pm$ 4.71 & -22.6 & -31.8 & -11.4 \cr
NGC2264 & 06 38 25.385 & 09 32 14.459 & 44.18 $\pm$ 2.21 & 60.55 $\pm$ 1.89 & 7.1 & 6.5 & 9.1 \cr
\noalign{\smallskip}
\hline
\end{tabular}
\end{flushleft}
\end{table}
\normalsize

%

%

\section{Preliminary Results on other YSOs} 

 In this section we present very preliminary results of a H$_2$O maser survey
 performed by us towards a sample of 5 massive YSOs similar to IRAS
 20126+4104 (see Table~1). All of our maser sources are associated with
 molecular outflows, have luminosities in excess of 2000~$L_\odot$, and are
 not associated with a detectable HII region (see Tofani et al. 1995).
 Although the data reduction is still in progress and only maps for the most
 intense spectral features have been produced so far, we may summarize the
 results obtained for two of the sources: NGC281-W and GGD 12-15.

\newpage

\noindent
\underline{{\it NGC281-W}} \\

\begin{figure}[h]
   \centering
   \vspace{250pt}
 \includegraphics{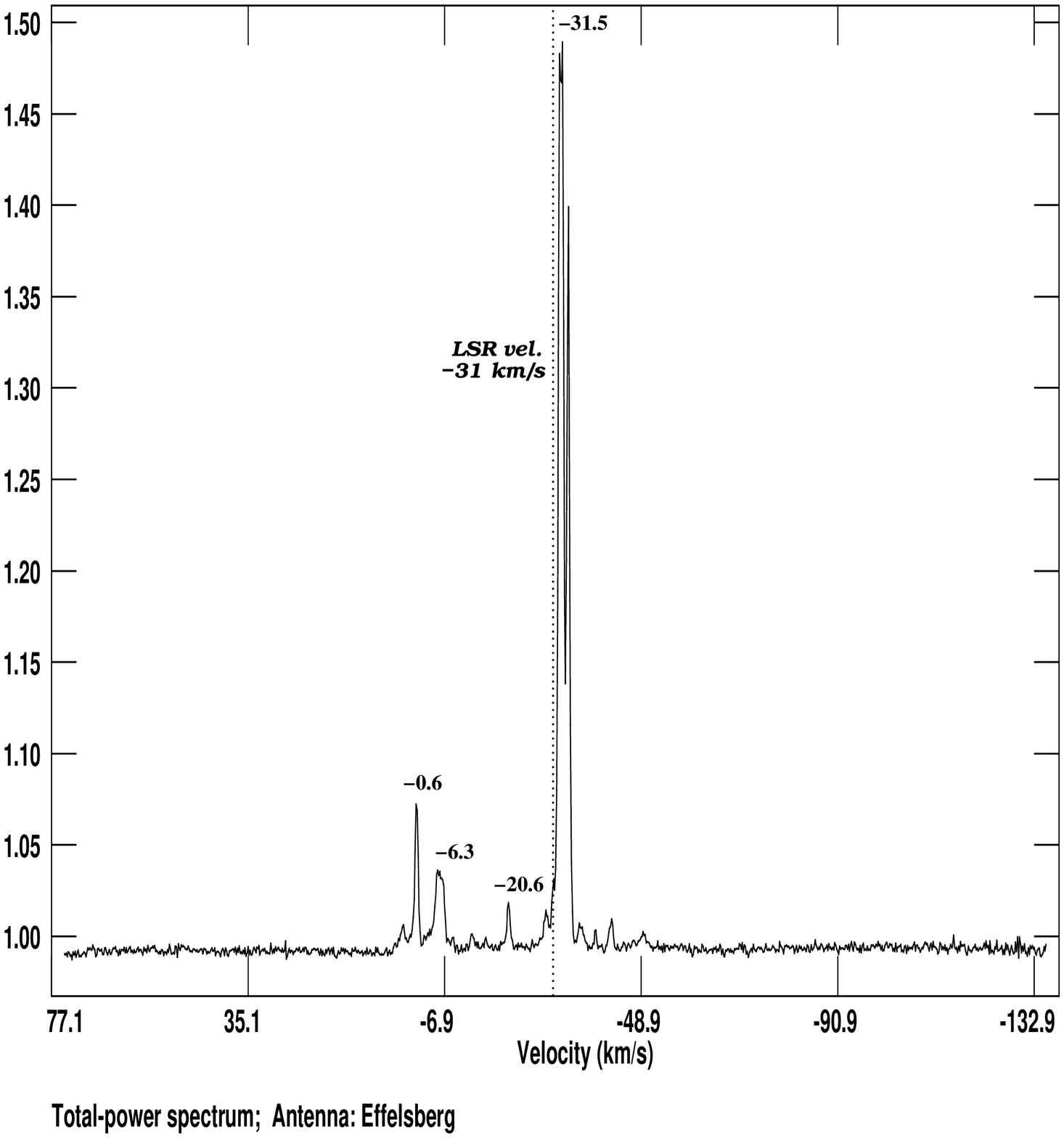}
 \includegraphics{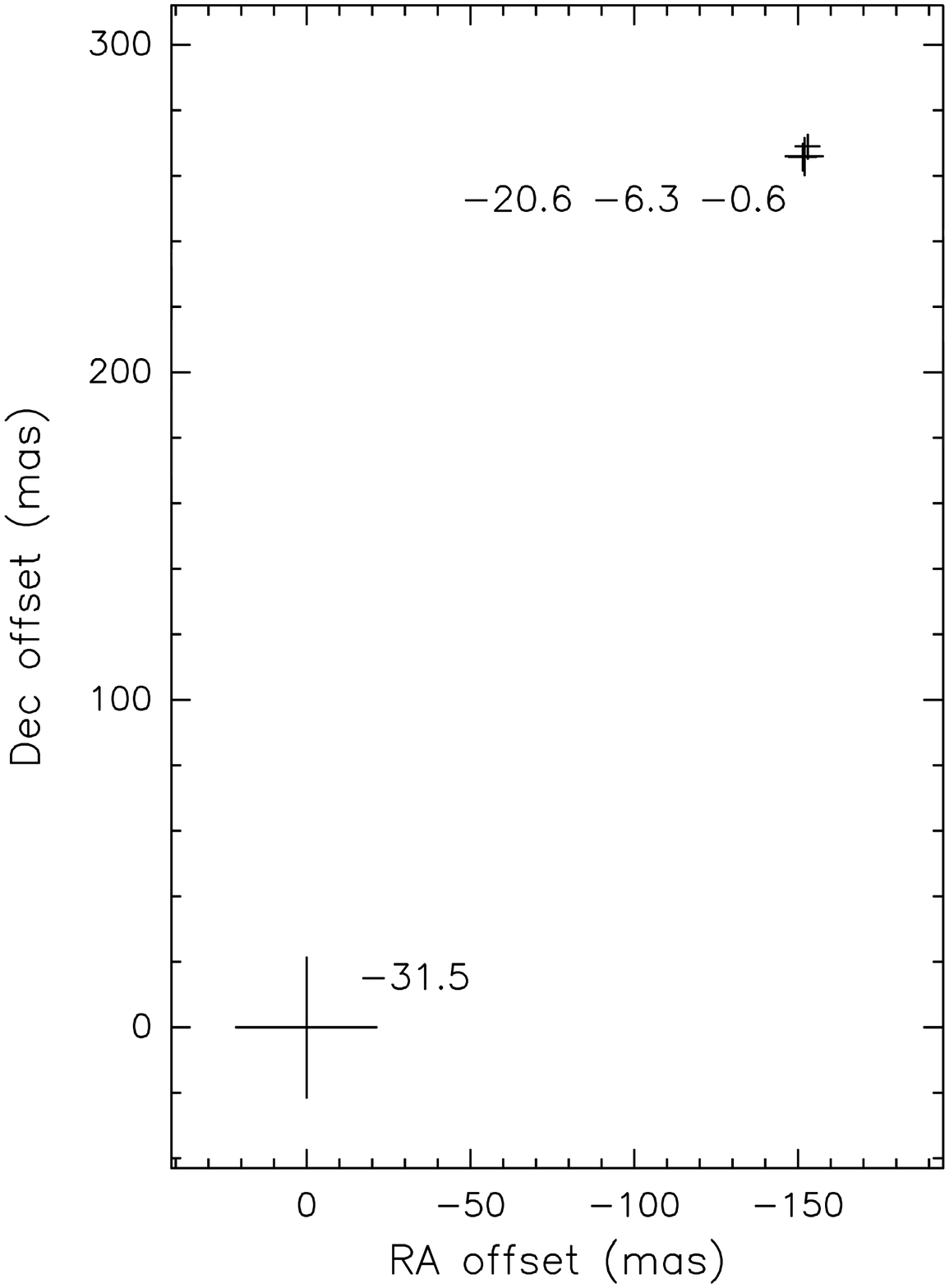}
\vspace{150pt}
      \caption{{\it Upper}: Effelsberg total power spectrum for NGC281-W.
The vertical dotted line indicates the systemic LSR velocity.
The labels indicate the velocities in [km/s]
of the most prominent features, for which we have made maps. 
{\it Lower}: Distribution of H$_2$O maser spots corresponding to the 
dominant emission features in the total power spectrum. The channel with 
the highest emission (vel. -31.5 km ~s$^{-1}$) was used as reference in 
the astrometric analysis. 
The labels in the plot indicate the velocities corresponding to each spot,
in [km/s]; the size of the crosses is proportional to the square root of 
the peak flux in the maps.
         \label{fig:ngc281w}
         }
   \end{figure}

Fig.~\ref{fig:ngc281w} ({\it upper}) shows the total power spectrum 
obtained with the Effelsberg radiotelescope. The systemic LSR velocity, 
derived from single dish observations is indicated with a dotted line. 
The labels on top of the most prominent features correspond to the velocities
of the channels, in units of [km s$^{-1}$], for which we have produced maps. 

The {\it lower} plot in Fig.~\ref{fig:ngc281w} shows the distribution of
H$_2$O maser spots corresponding to the most prominent features in the total
spectrum shown above.
The location of the spots is given by the crosses, whose size is proportional
to the square root of the peak flux in the individual maps for the velocities
indicated aside. The channel containing the most prominent peak 
of emission (velocity -31.5 km s$^{-1}$) was selected as reference in the
analysis, and corresponds to the spot with null righ ascension and declination
offsets in the map. 
The other spots correspond to emission from secondary peaks in the total 
spectrum, with red-shifted emission with respect to the systemic LSR velocity 
of the bulk material (velocities -6.3,-0.6 and -20.6 km s$^{-1}$). 
The spots in the map appear aligned along NW-SE direction. This direction 
does not seem to coincide with that from larger scale structures seen in 
observations of other outflow tracers (Cesaroni et al. 1999b, and references 
herein). On the other hand, large scale structures can also be affected by the 
presence of multiple outflows arising from a common star forming region. \\

%

\vspace*{0.75cm}

\noindent
\underline{{\it GGD 12-15}} \\

\begin{figure}
   \centering
   \vspace{500pt}
 \includegraphics{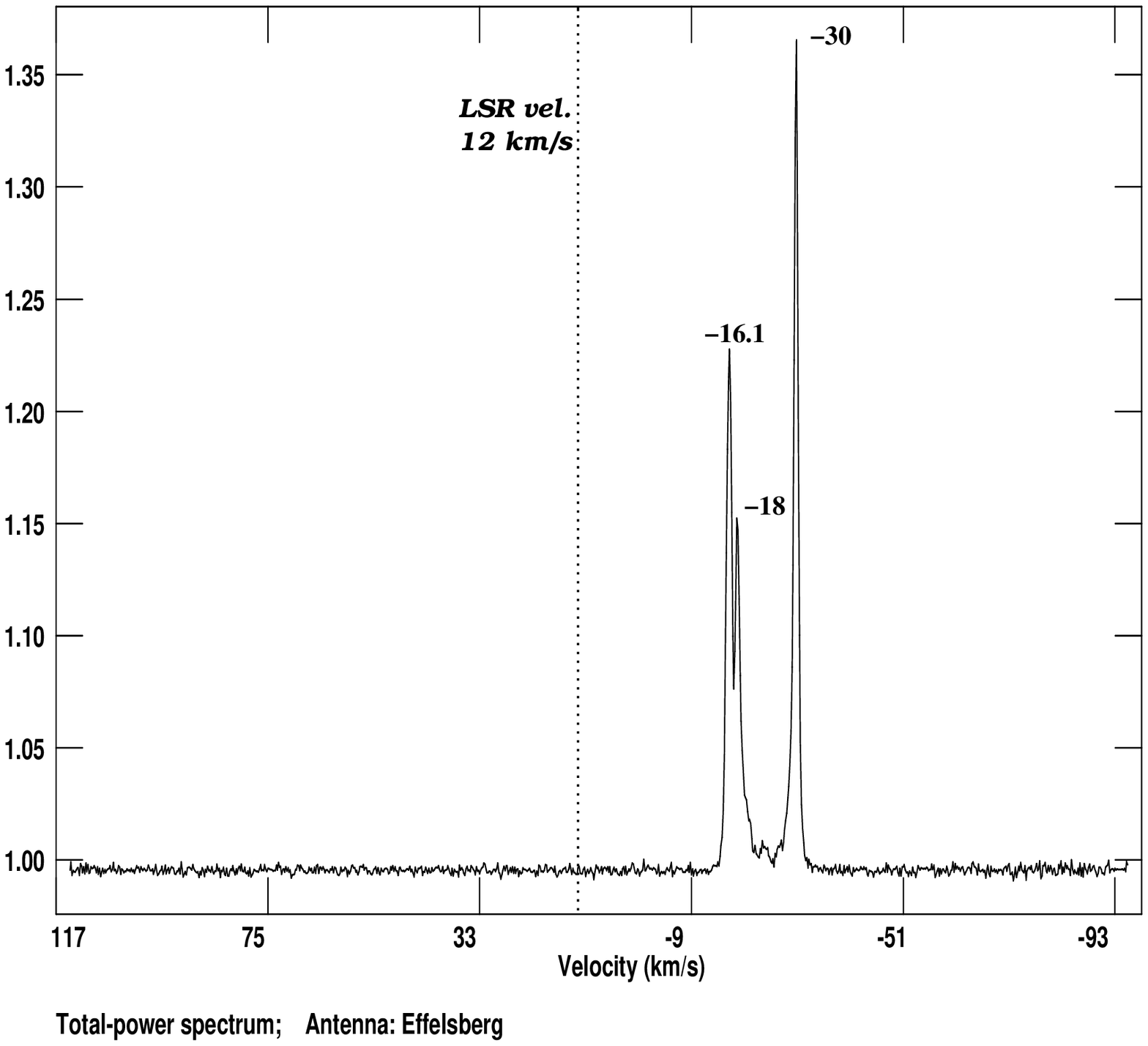}
 \includegraphics{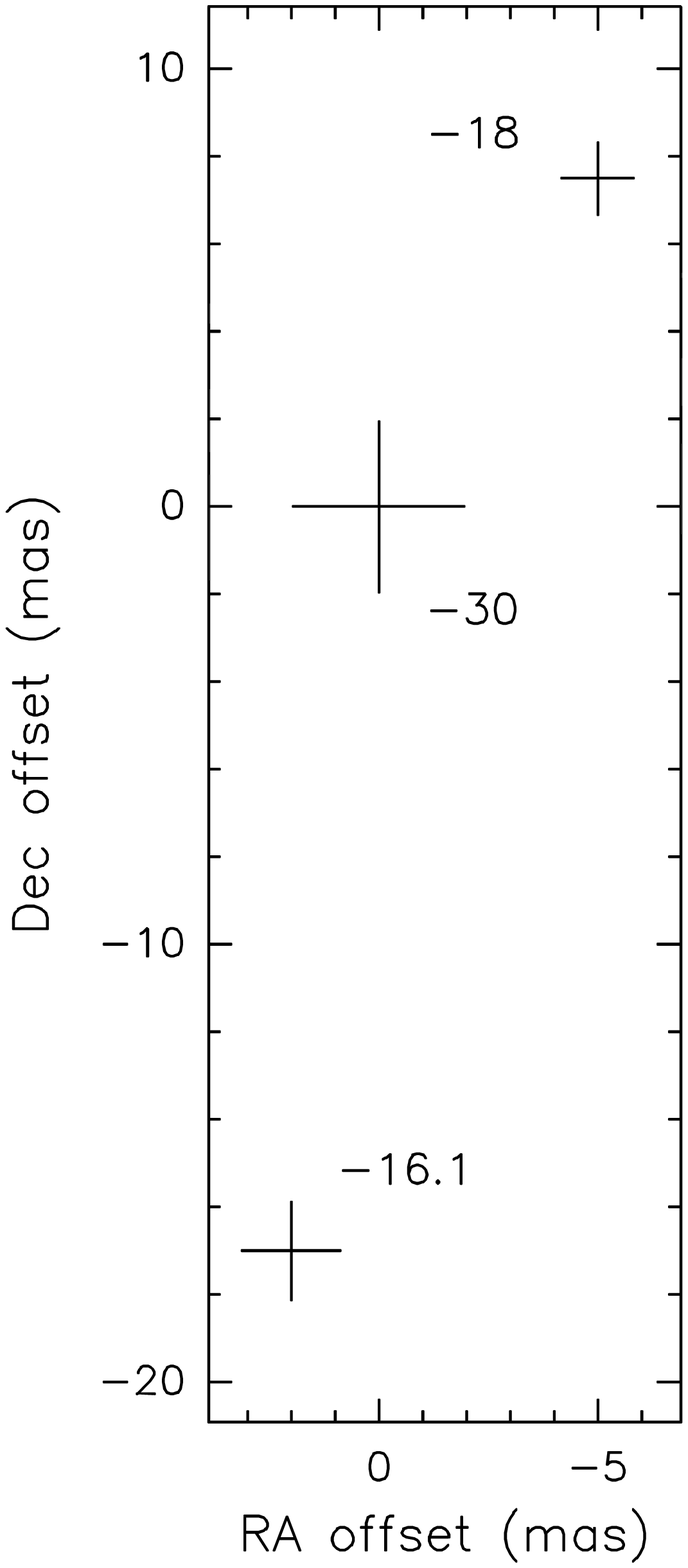}
      \caption{{\it Upper}: Effelsberg total power spectrum for GGD12-15.
The vertical dotted line indicates the systemic LSR velocity.
The labels indicate the velocities in [km/s]
of the most prominent features, for which we have made maps. 
{\it Lower}: Distribution of H$_2$O maser spots corresponding to the 
dominant emission features in the total power spectrum. The channel with 
the highest emission (vel. -30 km ~s$^{-1}$) was used as reference in 
the astrometric analysis. 
The labels in the plot indicate the velocities corresponding to each spot,
in [km/s]; the size of the crosses is proportional to the square root of 
the peak flux in the maps.
         \label{fig:ggd1215}
         }
   \end{figure}

Fig.~\ref{fig:ggd1215} ({\it upper}) shows the total power spectrum 
obtained with the Effelsberg radiotelescope, along with the systemic LSR
velocity obtained from single dish observations (dotted line). Also,
it includes labels indicating the velocity of the dominant features 
in the spectrum, for which we have produced maps.

The {\it lower} plot in Fig.~\ref{fig:ggd1215} shows the distribution of
H$_2$O maser spots corresponding to the most prominent features in the total
spectrum shown above. The sizes of the crosses, and the 
labels in the plot have the same meaning as explained for the other source. 
In the analysis, the channel with the most prominent peak of emission 
(velocity -30 km s$^{-1}$) was selected as reference. 
The spots corresponding to secondary peaks of emission in the total spectrum
are located on opposite sides of the reference maser feature, but
the fact that all spots are blue-shifted does not seem to support an 
association of the H$_2$O maser emission with a bipolar outflow, 
although this cannot be ruled out  (see the case of IRAS20126+4104). 
In fact, the (NW-SE) direction of H$_2$O maser emission in our phase 
referenced maps is in agreement with that of J=2-1 CO line emission from
the bipolar outflow detected by Little et al. 1990.

%

\section{Conclusions}

We have used global VLBI multiepoch observations to study the H$_2$O maser 
emission towards IRAS 20126+4104, using phase referencing techniques. 
The map of the H$_2$O maser spots, along with the measured proper motions,
prove that the water masers are clearly tracing a well collimated outflow. 
Hence, are excellent tools to investigate the structure and kinematics of
molecular jet/outflows in massive YSO.

We aim to extend the observations to a larger sample of protostars, selected
from the Tofani et al. (1995) catalog.
With that in mind, we have observed a first epoch EVN observations of a 
selection of 5 YSOs, and preliminary results have been presented in that
paper. We must wait for a more detailed analysis before drawing any
conclusion.

\cleardoublepage

\end{document}